\documentclass{article}
\usepackage{bm}
\usepackage{indentfirst}
\usepackage{natbib}
\usepackage{graphics}
\usepackage{graphicx}
\usepackage{amsmath}
\usepackage{verbatim}
\usepackage{multirow}
\usepackage{float}
\usepackage{tikz}
\usepackage{amsmath, nccmath}
\usepackage{geometry}
\usepackage[colorlinks=true,linkcolor=blue,urlcolor=black,bookmarksopen=true,citecolor=blue]{hyperref}
\usepackage{bookmark}
\usetikzlibrary{decorations.pathreplacing}
\usepackage{setspace}
\usepackage{comment}
\usepackage{adjustbox}
\usepackage{xcolor}
\usepackage{chngcntr}
\usepackage{authblk}
\usepackage{algorithm}
\usepackage{algpseudocode}
\usepackage{caption}

\newcounter{algsubstate}
\renewcommand{\thealgsubstate}{\alph{algsubstate}}
\newenvironment{algsubstates}
{\setcounter{algsubstate}{0}%
	\renewcommand{\State}{%
		\stepcounter{algsubstate}%
		\Statex {\footnotesize\thealgsubstate:}\space}}
{}

\begin{document}
\title{A Bayesian Hidden Semi-Markov Model with Covariate-Dependent State Duration Parameters for High-Frequency Data from Wearable Devices}
\date{}
\author[1]{Shirley Rojas-Salazar}
\author[1]{Erin M. Schliep}
\author[1]{Christopher K. Wikle}
\author[2]{Matthew Hawkey}
\affil[1]{Department of Statistics, University of Missouri}
\affil[2]{Institute for Health and Sport, Victoria University}

\maketitle

\begin{abstract}
Data collected by wearable devices in sports provide valuable information about an athlete's behavior such as their activity, performance, and ability. These time series data can be studied with approaches such as hidden Markov and semi-Markov models (HMM and HSMM) for varied purposes including activity recognition and event detection. HSMMs extend the HMM by explicitly modeling the time spent in each state. In a discrete-time HSMM, the duration in each state can be modeled with a zero-truncated Poisson distribution, where the duration parameter may be state-specific but constant in time. We extend the HSMM by allowing the state-specific duration parameters to vary in time and model them as a function of known covariates derived from the wearable device and observed over a period of time leading up to a state transition. 
In addition, we propose a data subsampling approach given that high-frequency data from wearable devices can violate the conditional independence assumption of the HSMM. We apply the model to wearable device data collected on a soccer referee in a Major League Soccer game. We model the referee's physiological response to the game demands and identify important time-varying effects of these demands associated with the duration in each state.
\end{abstract}
Keywords: HSMM; MCMC; data subsampling; activity recognition; soccer referee.

\doublespacing
\section{Introduction}
\label{sec:int}

Wearable devices are accessories, or smart clothes, worn on or near the body \citep{motti_introduction_2020} that provide users with information regarding, among other things, their health \citep{wu_materials_2018} and  physical activity \citep{li_wearable_2016}. The use of this technology in sports \citep{kos_wearable_2017} provides information to athletes, trainers, coaches, and physicians, and can be used to assess the physical demands of training \citep{seshadri_wearable_2017}, evaluate performance \citep{bachlin_swimming_2012}, and prevent or detect injuries \citep{li_wearable_2016}. The data from wearable sports devices are collected at an extremely high-frequency for a large number of variables, including acceleration, velocity, and heart rate \cite[]{kos_wearable_2017,motti_introduction_2020}.

Time series data collected by wearable devices can be studied using a wide range of analyses, depending on the objectives, resulting in varied inference. One important application of wearable device data is activity recognition. \cite{sztyler_-body_2016} investigated if the positioning of a wearable device on a person's body impacts activity recognition. They developed a classifier that determined the on-body position of the wearable device, compared activity classifiers that were position-independent and position-aware, and determined the best method was a position-aware random forest classifier. In another application, \cite{kos_tennis_2018} created a tennis stroke database using miniature wearable devices and video data to assess stroke consistency. \cite{chereshnev_gain_2018} described a system for human gait analysis, where methods such as dynamic Bayesian networks and recurrent neural networks were applied to data from wearable body sensors to predict how amputated legs move, and to help in the control of robotic leg prostheses.

Hidden Markov and hidden semi-Markov models provide an alternative approach for analyzing time-series data from wearable devices. A hidden Markov model (HMM) consists of a sequence of unobserved discrete states and another set of observable random variables that are assumed conditionally independent given the state at each observed time point \citep{rabiner_tutorial_1989}. The transition from one state to another depends on a transition probability, which is defined conditionally on the current state, and where the probability of self-transitioning (i.e., remaining in the same state) is non-zero. This non-zero probability implies that the time spent in each state follows a geometric distribution \citep{yu_hidden_2010}. However, this distributional assumption may not be realistic for some processes, making it necessary to additionally model the state duration. This model extension defines the hidden semi-Markov model (HSMM) \citep{yu_hidden_2016}.

Different applications of HMMs and HSMMs can be found in sports, with some utilizing data from wearable devices. In particular, \cite{motoi_bayesian_2012} used a Bayesian HMM to create metadata for sports games through event detection, and their method was evaluated using video data of soccer games to detect events, such as kick off, corner kick, or goal kick. \cite{xie_structure_2004} applied the HMM  to two defined states of a soccer game, play and break, using the features dominant color ratio and motion intensity that are obtained from video data. Similar types of data have been analyzed with HSMMs. For example, \cite{itoda_model-based_2015} analyzed handball data (a goal-type game like soccer) and used HSMMs in their analysis to segment the behaviors of the players. The information from the segments was then used to extract the causality in the behavior sequences among the players. Specifically, they used a metric of the information shared between pairs of behavior sequences at multiple time delays to investigate how the individual behaviors of the players change based on the behavior of other players.  \cite{thomas_wearable_2010} performed activity recognition using both an HMM and a semi-Markov model (SMM) applied to swimming data collected with a wearable sensor, where the segmentation of the session improved comprehension of the training. 

When the duration in an HSMM is modeled with a Poisson distribution, the duration parameter, which can be different for each hidden state, is assumed to be constant in time. This assumption, however, might not be reasonable in all cases. For example, if we consider a cyclist in a race, with states generally representing low and high intensity, the amount of time spent in each of the states could be related to environmental variables such as wind direction and elevation gradient, as well as changes in the cyclist's physical state throughout the race. In the context of soccer, the average duration of a player in a state might be longer dependent on field location, during a long period where the ball leaves the field of play, or the last few minutes of a close scoring game. Under both situations, it becomes necessary to extend the HSMM to allow the duration parameters to  depend on covariates and vary in time.

By extending the HSMM to incorporate covariates in the duration parameter, we can identify the factors associated with the time spent in the different states. For example, when there is a state transition, the duration parameter for the new state could be modeled as a function of covariates observed in the period leading to the transition, or the value of the covariate at the moment right before the switch. The functional relationship between covariates and the parameters of the duration distribution could be state-specific, and modeling these relationships can provide important inference with regard to their extent and direction.
Importantly, the inference is not obtained at the high-frequency level at which the data are collected, rather it is obtained in terms of the duration intervals.

In both HMMs and HSMMs, observations are described with an emission distribution and are assumed to be conditionally independent given the state, meaning they are independent of previous states and observations \cite[]{pohle_selecting_2017,yu_hidden_2016}. However, high-frequency data, such as data obtained from wearable devices, are more likely to be correlated. The violation of the conditional independence assumption can have dramatic impacts on statistical inference.  \cite{pohle_selecting_2017} presented a simulation study to determine the effects of assumption violations in the selection of the number of states in HMMs, and determined that when the conditional independence assumption is violated, and the Akaike and Bayesian information criteria are used to select between models, the number of states will be overestimated. 

Markov-switching regression models (MSR) and neural networks (NN) are two approaches that have been used when the conditional independence assumption is not met. In MSR, the observations are modeled as a function of covariates or as an autoregressive model \citep{langrock_markov-switching_2017}. In the context of NN, \cite{ravuri_how_2016} apply a deep neural network HMM (DNN-HMM) and show that deeper NNs compensate for the conditional independence assumption violation more than shallow NNs. In addition, \cite{dai_recurrent_2017} consider a recurrent hidden semi-Markov model (R-HSMM) that incorporates a recurrent neural network (RNN) in the observation model of an HSMM to accommodate more complex dependencies in the observation sequence.

The disadvantage of approaches such as NN is that they are computationally expensive to implement. Thus, it is useful to consider a more computationally tractable approach for mitigating the conditional dependence. One approach that has not been considered in this context is data subsampling. Data subsampling is used for reducing computational cost or in determining the sampling distribution of a statistic. For instance, it has been used as an alternative to deal with large datasets to increase efficiency in Firefly Monte Carlo (FlyMC) \citep{maclaurin_firefly_2015} and in subsampling Markov chain Monte Carlo (MCMC) \citep{quiroz_speeding_2019}. Both of these approaches present an MCMC sampling algorithm that considers only a subset of the data at each iteration. Several experiments were conducted to evaluate the performance of FlyMC and they found it to be more efficient than standard MCMC sampling in many instances. Similarly, \cite{quiroz_speeding_2019} showed that subsampling MCMC is often more efficient than standard MCMC and other competing subsampling algorithms. These studies suggest that, for an HSMM, random data subsampling can be introduced as part of the MCMC algorithm as an attempt to reduce or eliminate the conditional dependence in the data.

The goal of this paper is to develop an HSMM with time-varying duration parameters that are dependent on covariates and apply it to data collected from wearable devices. Specifically, we model high-frequency heart rate data of a soccer referee obtained from a wearable device to determine the state sequence of the referee's physiological response to movement during a game. We use covariates to explain the variation in the duration in each state and obtain inference on important characteristics. Previous approaches using HMMs or HSMMs have included covariates in the observation model or in the specification of transition probabilities  \cite[e.g.][]{koki_forecasting_2020,economou_mcmc_2014,titman_semi-markov_2010}, but the inclusion of covariates in the model for state durations has not been considered. Additionally, we propose a novel data subsampling approach to mitigate the violation of the conditional independence assumption that is common in high frequency data.

In Section \ref{sec:data} we present the soccer referee data used in this analysis. Section \ref{sec:model} provides a description of the HSMM, which defines the duration distribution parameter as a function of covariates, as well as the details of a simulation study to investigate the effects of dependence in the observation sequence. The results of the model applied to the referee dataset are presented in Section \ref{sec:appl}. Lastly, Section \ref{sec:disc} provides a discussion and conclusion, as well as future extensions. 

\section{Motivating tracking data}
\label{sec:data}
The wearable device data used in this application come from high-resolution tracking data of a Major League Soccer (MLS) referee in a game from 2017. The data were recorded with a minimax device from Catapult and recovered through their Sprint software \citep{sprint_2013}. The raw data were collected at a rate of 10Hz for approximately 150 minutes, starting at the warm-up and extending through the end of the game. The data were truncated for this analysis to the 90 minutes of the game and were thinned to 1Hz (5400 time points). The data extracted from the wearable device worn by the referee consisted of nearly sixty variables, including measures of movement (velocity and acceleration), global positioning information (i.e., latitude, longitude, and altitude), and physio-chemical variables (heart rate, metabolic power, and player load).

\begin{figure}[H]
	\begin{center}
		\includegraphics[scale=.55]{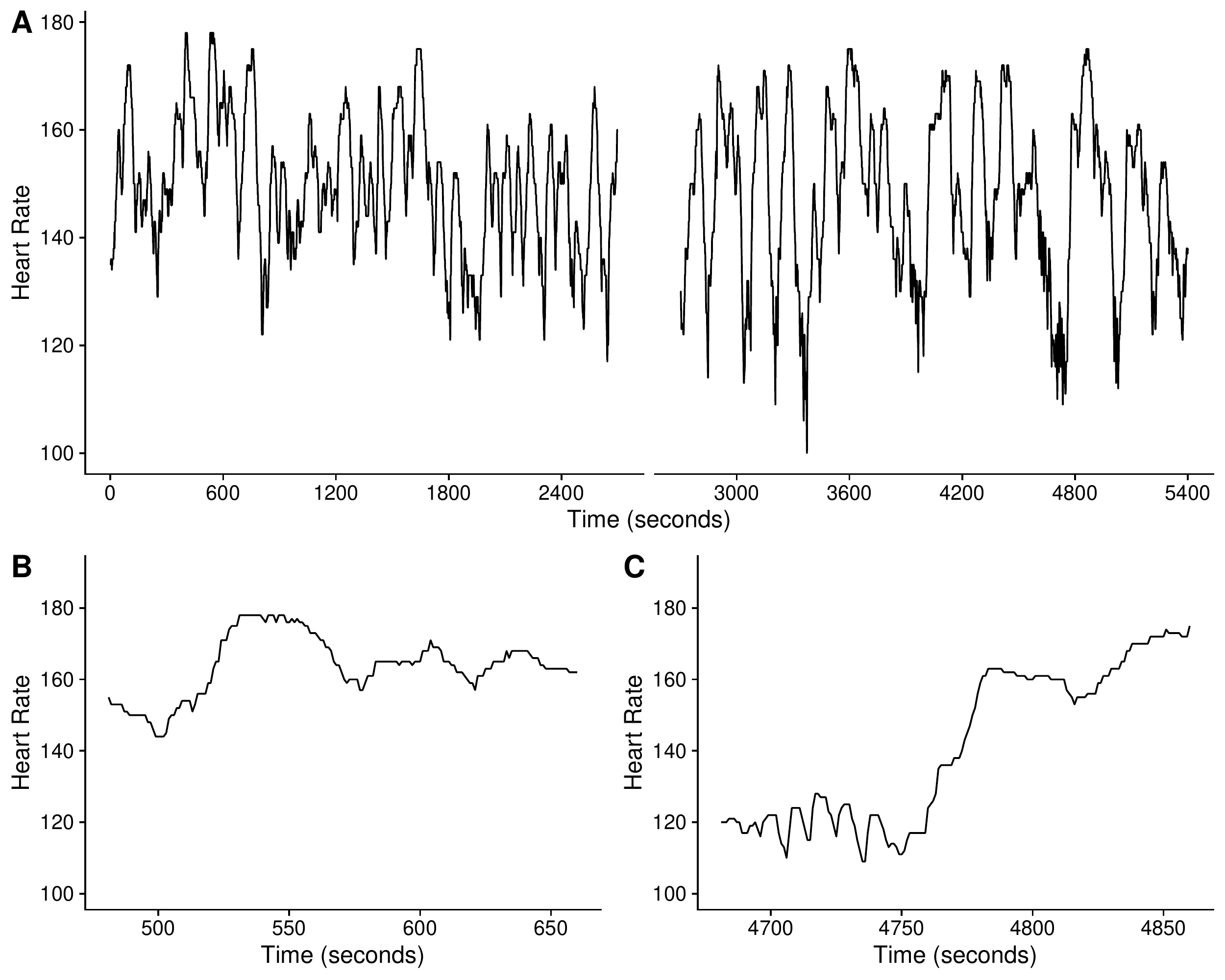}
		\caption{Referee heart rate (bpm) during the game. Panel \textbf{A}. Full game. Panels \textbf{B} and \textbf{C} show zoomed in 3-minute periods in the first and second half respectively, \label{FigureG3}}
	\end{center}
\end{figure}

The objective of our analysis is to describe the latent states associated with the referee's physiological response to the demands of the game. As such, we consider the instantaneous heart rate (HR) in beats per minute (bpm) as our observation sequence (Figure \ref{FigureG3}). Heart rate measured during the game provides information about the exigency on the physiology of the referee \citep{cerqueira_analysis_2011} and can help inform the state of activity or intensity of the referee throughout the game. The average heart rate during the 90 minute game for this referee was 150bpm. This is consistent with the results presented in \cite{cerqueira_analysis_2011}, who reported a range of average heart rates between 141 to 165 for soccer referees in different competitions of Fédération Internationale de Football Association (FIFA). Figure \ref{FigureG3} (panel A) shows the heart rate during the game, whereas panels B and C show a more detailed portion of the HR measurements from minute 8 to 11 of the first half, and minute 33 to 36 of the second half, respectively.

The variables we consider as possible covariates to capture the time variation in the state durations are acceleration, cumulative distance traveled, and distance from the center of the field. According to the Sprint manual \citep{sprint_2013}, Doppler based GPS acceleration is the rate of change in velocity, and the odometer indicates the distance traveled in meters. The variable distance from the center was computed as the distance in meters from the referee's location to the center of the field, noting that referees move in a diagonal trajectory across the field. The average acceleration during the game was 0.03$m/s^2$ with a range of -5.97 to 8.08$m/s^2$. The total distance traveled was approximately 10$km$, and the distance from the center ranged from 0.4 to 47.4$m$, with a mean of 19$m$.

\section{Hidden semi-Markov model with covariate-dependent duration parameters}
\label{sec:model}
We begin by specifying the HMM and important notation. Then, we extend the HMM to the HSMM, develop the state-specific duration model as a function of covariates, and describe the methods for Bayesian inference. Lastly, we propose a simulation to investigate the implications of the violation of the conditional independence assumption.

\subsection{Hidden Markov model and notation}
\label{subsec:HMM}

In a discrete-time HMM, the sequence of observations from time $1$ to $n$ can be denoted as $\mathbf{y}= (y_{1}, \dots,y_{n}) ^\prime$. The corresponding sequence of unobserved states is denoted as $\mathbf{S}= ( S_{1},\dots,S_{n}) ^\prime$, where $S_{i} \in \left\lbrace 1,2, \dots, M \right\rbrace $, $i=1,\dots,n$, and $M$ is the total number of unique states. The state at time $1$ has a distribution defined by $\rho_j=p\left[ S_{1}=j\right]$, $j=1,\dots,M$. The transition to the next state, $S_{2}$, is conditional on the state $S_{1}$ according to the Markov property. In general, the transition probability matrix $\bf{P}$ provides the probabilities of transitioning from one state to another when the state space is discrete and constant in time. The matrix $\bf{P}$, has entries $p_{j,k}$, with $p_{j,k}=p\left[ S_{i+1}=k \mid S_{i}=j \right]$, where $1 \leq j,k \leq M$, and $\sum_{k=1}^{M} p_{jk} =1 $.

Observations are emitted by each of the states in the hidden sequence (Figure \ref{Figurediagram}A) following a state-dependent probability distribution $f(\mathbf{y}|\bm{\theta},\mathbf{S})$. Assuming the observation distribution is Gaussian, the parameters $\bm{\theta}$ correspond to the mean and variance for each state: $\bm{\mu}=(\mu_1,\cdots,\mu_M)^\prime$ and $\bm{\sigma}^2=(\sigma^2_1,\cdots,\sigma^2_M)^\prime$. The joint likelihood of the observations can be written as:

\begin{equation*}
L(\mathbf{y} \mid \bm{\mu}, \bm{\sigma}^2, \mathbf{S})=\prod_{i=1}^{n} f(y_{i}|\mu_{S_{i}},\sigma_{S_{i}}^2),
\end{equation*}

\noindent and the likelihood of the Markov chain is:
\begin{equation*}
L(\mathbf{S} \mid \bm{\rho}, \mathbf{P})= \rho_{S_{1}}  \cdot \prod\limits_{i=1}^{n-1} p_{S_{i},S_{{i+1}}}.
\end{equation*} 

The complete likelihood of the HMM is the joint likelihood of observations and states:
$L(\mathbf{y}, \mathbf{S} \mid \bm{\mu}, \bm{\sigma}^2, \bm{\rho}, \mathbf{P}) = L(\mathbf{y} \mid \bm{\mu}, \bm{\sigma}^2, \mathbf{S}) \times L(\mathbf{S} \mid \bm{\rho}, \mathbf{P})$. In summary, an HMM with $M$ states and $n$ observations has a set of model parameters  that includes the emission distribution parameters $\bm{\mu}$ and $\bm{\sigma}^2$, the initial distribution probabilities $\bm{\rho}$, and the transition probability matrix $\mathbf{P}$.

\bigskip

\begin{figure}[H]
	\centering
	
	\begin{tikzpicture}
	
	\node at (-1.0, 1.0) {\textbf{A}};
	
	\node at (0.0, 0.0) {\textit{States}};
	\node at (0.0, -1.0) {\textit{Observations}};
	
	\node (St1) at (1.8, 0.0) {$S_{{1}}$};
	\node (ys1) at (1.8, -1.0) {$y_{{1}}$};
	\draw[->] (St1) to (ys1);
	\node (St2) at (3.8, 0.0) {$S_{{2}}$};
	\node (ys2) at (3.8, -1.0) {$y_{{2}}$};
	\draw[->] (St2) to (ys2);
	
	\node (St3) at (5.8, 0.0) {$S_{{3}}$};
	\node (ys3) at (5.8, -1.0) {$y_{{3}}$};
	\draw[->] (St3) to (ys3);
	
	\draw[->] (St1) to (St2);
	\draw[->] (St2) to (St3);
	\node (dotss) at (6.8, 0.0) {$\cdots$};
	\node (dotsy) at (6.8, -1.0) {$\cdots$};
	
	\node (Stn) at (7.8, 0.0) {$S_{{n}}$};
	\node (ysn) at (7.8, -1.0) {$y_{{n}}$};
	\draw[->] (Stn) to (ysn);
	
	
	\node at (-1.0, -2.0) {\textbf{B}};
	
	\node at (0.0, -3.0) {\textit{States}};
	\node (S1) at (1.8, -3.0) {$S_{1}$};
	\node (S2) at (4.7, -3.0) {$S_{2}$};
	\node (S3) at (9.6, -3.0) {$S_{Q}$};
	
	\draw[->] (S1) to (S2);
	\node (dotss) at (8.4, -3.0) {$\cdots$};
	\node (dotsy) at (8.4, -4.0) {$\cdots$};
	
	\node at (0.0, -4.0) {\textit{Observations}};
	
	\node (y11) at (1.8, -4.0) {$y_{{1}}$};
	\node (y12) at (2.4, -4.0) {$y_{{2}}$};
	\node (dots11) at (2.9, -4.0) {$\cdots$};
	\node (yd1) at (3.5, -4.0) {$y_{{\tau_1}}$};
	\draw[->] (S1) to (y11);
	\draw[->] (S1) to (y12);
	\draw[->] (S1) to (yd1);
	
	\node (y21) at (4.7, -4.0) {$y_{{T_1+1}}$};
	\node (y22) at (5.8, -4.0) {$y_{{T_1+2}}$};
	\node (dots2) at (6.6, -4.0) {$\cdots$};
	\node (yd2) at (7.5, -4.0) {$y_{{T_1+\tau_2}}$};
	\draw[->] (S2) to (y21);
	\draw[->] (S2) to (y22);
	\draw[->] (S2) to (yd2);
	
	\node (y31) at (9.6, -4.0) {$y_{{T_{Q-1}+1}}$};
	\node (y32) at (11.0, -4.0) {$y_{{T_{Q-1}+2}}$};
	\node (dots3) at (12.05, -4.0) {$\cdots$};
	\node (yd3) at (13.1, -4.0) {$y_{{T_{Q-1}+\tau_Q}}$};
	\draw[->] (S3) to (y31);
	\draw[->] (S3) to (y32);
	\draw[->] (S3) to (yd3);
	
	\node at (0.0, -5.2) {\textit{Durations}};
	\draw [decorate,decoration={brace,raise=2mm,amplitude=3pt,mirror}] (y11.south west) -- (yd1.south east);
	\draw [decorate,decoration={brace,raise=2mm,amplitude=3pt,mirror}] (y21.south west) -- (yd2.south east);
	\draw [decorate,decoration={brace,raise=2mm,amplitude=3pt,mirror}] (y31.south west) -- (yd3.south east);
	\node (tau1) at (2.8, -5.2) {$\tau_1$};
	\node (tau2) at (6.2, -5.2) {$\tau_2$};
	\node (tau3) at (11.5, -5.2) {$\tau_Q$};
	
	\node (dotss) at (8.4, -5.2) {$\cdots$};

	\end{tikzpicture}
	
	\caption{State and observation sequences. Panel \textbf{A}. HMM: One observation is emitted by each state in the sequence. Panel \textbf{B}. HSMM: Several observations are emitted by each state, the number is determined by the duration in the state.} \label{Figurediagram}
\end{figure}

\subsection{Hidden semi-Markov model}
\label{subsec:HSMM}

Figure \ref{Figurediagram}B illustrates the HSMM where instead of assuming there is only one observation per state, a sequence of observations are emitted. The number of observations depends on the amount of time spent in the state. Following the notation in \cite{economou_mcmc_2014}, $\tau$ represents the length of time that the sequence remains in a state before transitioning. These \emph{durations} are labeled in Figure \ref{Figurediagram}B as $\tau_1, \dots, \tau_Q$, where Q is the number of intervals or segments. 
Letting $q=1, \dots, Q$ we define $T_q$ to denote the cumulative duration in segments $1$ through $q$.
Lastly, we define $h_j(\tau \mid \phi_j)$ as the duration distribution for each state $j$, $j=1,\dots,M$, with parameter $\phi_j$. 

Similar to the HMM, the likelihood of the HSMM has two main components consisting of the likelihood of the observations conditional on the states and the likelihood of the semi-Markov chain of states. The joint likelihood of the observations can be specified analogous to the HMM case, but is written incorporating the segment-specific notation: 

\begin{equation}
L(\mathbf{y}|\bm{\mu}, \bm{\sigma}^2, \mathbf{S})= \prod_{i=1}^{n} f(y_{i}|\mu_{S_{i}},\sigma_{S_{i}}^2)= \prod_{q=1}^{Q} f(\mathbf{y}_{\tau_q}|\mu_{S_{q}},\sigma_{S_{q}}^2), 
\end{equation}

\noindent where $\mathbf{y}_{\tau_q}$ corresponds to the vector of all the observations in time interval $q$. The likelihood of the state sequence includes the distribution of the first state, the transition probabilities for the state switches, as well as the information from the duration times:

\begin{equation}
\label{LMC}
L(S_1,\dots,S_Q, \tau_1,\dots,\tau_Q | \bm{\rho}, \mathbf{P},  \bm{\phi})= \rho_{S_1} \cdot \prod\limits_{q=1}^{Q-1}  h_{S_q}(\tau_q \mid \phi_{S_q}) \cdot p_{S_q,S_{q+1}}   \cdot h_{S_Q}(\tau_Q \mid \phi_{S_Q}).
\end{equation} 

\noindent Thus, the complete likelihood of the hidden semi-Markov model can be written as:

\bigskip

\hspace{2mm} $L(\mathbf{y}_{\tau_1},\dots,\mathbf{y}_{\tau_Q}, S_1,\dots,S_Q, \tau_1,\dots,\tau_Q \mid  \bm{\mu}, \bm{\sigma}^2,\bm\rho,\mathbf{P},\bm{\mathbf\phi})$

\begin{equation}
\label{hsmmlik}
\begin{gathered}
\hspace{50mm} =\rho_{S_1} \cdot \prod\limits_{q=1}^{Q-1}  h_{S_q}(\tau_q |\phi_{S_q}) \cdot p_{S_q,S_{q+1}} \cdot f(\mathbf{y}_{\tau_q}|\mu_{S_{q}},\sigma_{S_{q}}^2)   \\
\hspace{28mm}  \times \hspace{2mm} h_{S_Q}(\tau_Q |\phi_{S_Q}) \cdot f(\mathbf{y}_{\tau_Q}|\mu_{S_{Q}},\sigma_{S_{Q}}^2).
\end{gathered}
\end{equation}

\noindent Note, we have added the duration distribution parameters of each state to the list of parameters of the HMM. Specifically, the set of model parameters of the HSMM presented includes $\bm{\mu}, \bm{\sigma}^2,\bm\rho,\mathbf{P}$, and $ \bm{\phi}$.

\subsection{Use of covariates to model duration}
\label{subsec:cov}

Previous approaches have specified non-homogeneous HMM and HSMMs by modeling the parameters of the emission distribution or the probabilities of transition using covariates. 
We propose introducing non-homogeneity in the HSMM duration by letting the parameters of the state duration distribution vary in time as a function of covariates. If we let the duration distribution be a zero-truncated Poisson, we can define the duration parameter of the interval $q+1$, $\phi_{S_{q+1}}$, as a function of the covariate measurements observed prior to the transition at $T_{q}+1$. Notice that this specification enables the duration parameter to be both state-specific and vary in time. 

Let $\mathbf{X}$ be an $n \times r$ covariate matrix with rows corresponding to times $1$ to $n$, where $r$ is the number of covariates. Let $\bm\beta_{S_{q+1}}$ be an $(r+1)$-dimensional coefficient vector for state $S_{q+1}$ (accounting for an intercept in the model). Then the duration parameter for interval $q+1$, which we denote as $\phi_{S_{q+1}}(\mathbf{X}_{1:{T_{q}}},\bm \beta_{S_{q+1}})$, is a function of the covariate values observed up to time point $T_{q}$ (the first $T_q$ rows of $\textbf{X}$), and state specific coefficients $\bm{\beta}_{S_{q+1}}$. Here, $\phi_{S_{q+1}}$ can take any functional form of the covariates as long as $\phi_{S_{q+1}} > 0$. For example, we can write the function as:

\begin{equation}
\label{phif}
	\phi_{S_{q+1}}(\mathbf{X}_{1:{T_{q}}},\bm \beta_{S_{q+1}}) = g \left( \beta_{0,S_{q+1}} 
	+ \beta_{1,S_{q+1}} \cdot  f_1 \left( \mathbf x_{1,1:{T_{q}}} \right)   
	+ \cdots
	+ \beta_{r,S_{q+1}} \cdot  f_r \left( \mathbf x_{r,1:{T_{q}}} \right) 
	\right), 
\end{equation}

\noindent where $g(\cdot)$ is a specified function that ensures $\phi_{S_{q+1}} > 0$, and $f_1(\cdot), \dots, f_r(\cdot)$ can be any function of the covariates observed from time 1 to the time previous to the transition, $T_{q}$. The likelihood, which now includes the state-specific duration parameter function, can be written as: 

\bigskip

\hspace{2mm}  $L(y_{\tau_1},\dots,y_{\tau_Q}, S_1,\dots,S_Q, \tau_1,\dots,\tau_Q \mid  \bm{\mu}, \bm{\sigma}^2,\bm\rho,\mathbf{P},\mathbf{B},\mathbf{X})$
\begin{equation}
	\label{eqcov}
	\begin{gathered}
		= \rho_{S_1} \cdot h_{S_1}\left( \tau_1 \mid \phi_{S_1}(\mathbf{X}_{0},\bm \beta_{S_1}) \right) \cdot f(\mathbf{y}_{\tau_1}|\mu_{S_{1}},\sigma_{S_{1}}^2) \\
		\hspace{23mm}\times \hspace{2mm} \prod\limits_{q=2}^{Q}  h_{S_q}\left( \tau_q \mid \phi_{S_q}(\mathbf{X}_{1:{T_{q-1}}},\bm \beta_{S_q}) \right)  \cdot p_{S_{q-1},S_{q}} \cdot f(\mathbf{y}_{\tau_q}|\mu_{S_{q}},\sigma_{S_{q}}^2),
	\end{gathered}
\end{equation}

\noindent where $\mathbf{X}_{0}$ are the initial values for the covariates, and $\mathbf{B}$ is the matrix of $\beta$-coefficients with M rows and the number of columns is the number of covariates, $r$, plus an intercept:

\begin{equation*}
	\mathbf{B} =
	\begin{pmatrix}
		\beta_{0,1} & \beta_{1,1} & \cdots & \beta_{r,1}\\
		\beta_{0,2} & \beta_{1,2} & \cdots & \beta_{r,2}\\
		\vdots  & \vdots  & \ddots & \vdots \\
		\beta_{0,M} & \beta_{1,M} & \cdots & \beta_{r,M}\\
	\end{pmatrix}.
\end{equation*}

\noindent That is, $\bm\beta^\prime_{S_q}$ is the row of $\mathbf{B}$ that corresponds to state $S_q$. For example, when the state in interval $q$ is 1, then $\bm\beta_{S_q=1} = \left( \beta_{0,1}, \beta_{1,1}, \dots , \beta_{r,1} \right)^\prime $.


\subsection{Estimation of model parameters}
\label{subsec:estim}
Model inference can be obtained in a Bayesian framework using Markov chain Monte Carlo (MCMC) and a Metropolis-within-Gibbs sampling algorithm (see Appendix \ref{sec_appsa} for the detailed sampling algorithm). To complete the model specification, we assign diffuse priors to the model parameters. The means of the emission distribution are assigned independent Normal priors,  $N\left(0, 10000 \right)$ and the variances are assigned inverse-Gamma priors, $IG\left(3,3 \right)$. The initial probabilities, as well as each of the rows of the transition matrix, have Dirichlet priors, $Dir(1,1,1)$ and $Dir(1,1)$, respectively. The coefficient parameters in the model for the state-specific durations are assumed to be independent and distributed as $N \left(0, 10000 \right)$.

The posterior distribution of the states, intervals, and parameters of the HSMM can be summarized as:

\begin{equation}
\begin{split}
p(\bm{S},\bm{\tau},\bm{\mu},\bm{\sigma}^2,\bm{\rho},\mathbf{P},\mathbf{B} \mid \mathbf{y},\mathbf{X}) &\propto 
p(\mathbf{y} \mid \bm{S},\bm{\mu},\bm{\sigma}^2) \times p(\bm{\tau} \mid \bm{S},\mathbf{B},\mathbf{X}) \times p(\bm{S} \mid \bm{\rho},\mathbf{P})  
\times p(\bm{\mu} \mid \bm{\theta}_\mu,\bm{\lambda}_\mu^2) \\
&\times p(\bm{\sigma}^2 \mid \bm{\theta}_{\sigma^2},\bm{\lambda}_{\sigma^2})
\times p(\mathbf{B} \mid \bm{\theta}_B, \bm{\lambda}_B^2)
\times p(\bm{\rho}\mid \bm{\theta}_\rho) \times p(\bm{P} \mid \bm{\theta}_P).
\end{split}
\end{equation}

\noindent The state means and variances, initial probabilities, and transition probabilities can be sampled from their full conditionals using a Gibbs update, but a Metropolis algorithm is needed for the duration distribution coefficients. 

\cite{economou_mcmc_2014} provide an MCMC implementation of the HSMM using a forward algorithm to estimate the parameters, which alleviates the need to sample the state sequence in the process. However, our model requires sampling the states in order to obtain inference on the parameters of the duration distributions. The state sequence in an HMM can be sampled with the Viterbi algorithm \citep{viterbi_error_1967}. An extension of the Viterbi algorithm for the HSMM was proposed by \cite{yu_hidden_2016}, which we use to sample the state sequence in each iteration of our MCMC.
Additionally, the MCMC algorithm has to be adjusted to account for the conditional dependency in the observations. This modification is the inclusion of a subsampling approach that is discussed in the next section.

\subsection{Violation of the conditional independence assumption}
\label{subsec:sim}
We consider a simulation study to investigate the effect of the conditional independence assumption violation in the emission distribution parameter estimation. We simulate dependent data using an AR(1) model under several scenarios. Then, we fit an HSMM model to each generated dataset using a data subsampling approach, where at each iteration of the MCMC algorithm, an independent and random subset of the data is selected according to a specified sampling rate.

The different scenarios included in the simulation study are defined according to sample size, AR(1) autocorrelation parameter, and number of states. Three different number of states are considered ($M=$ 2, 3 and 4), while the possible values for the autocorrelation parameters are 0.3, 0.6, 0.86 and 0.95, as well as a case with no autocorrelation. Although the number of simulated observations varies across all realizations, we consider three cases consisting of approximately 1000, 2600 and 4000 observations. Overall, 45 scenarios were considered, each with 100 realizations. 

The procedure for simulating the data is as follows. First, the initial state, $S_1$, is sampled from its distribution. Second, the first duration, $\tau_1$, is generated from the corresponding zero-truncated Poisson distribution. The observations, $y_{1},y_{2},\cdots,y_{\tau_1}$, are generated from a Normal distribution with a specified autocorrelation. Then, the second state, $S_2$, is sampled conditional on $S_1$ according to the transition probability matrix $\textbf{P}$. After that, the duration is sampled as well as the observations emitted by that second state. The process continues until the specified sample size is reached.

Each of the simulated realizations is modeled with an HSMM assuming independence in the observations. For ease of computation, the states and other parameters are fixed and only the emission distribution parameters are estimated.

We investigate the benefit of the subsampling approach under various sample sizes, autocorrelation parameters, and number of states. In this approach, each iteration of the MCMC algorithm uses a different random subset of data according to a pre-specified percentage. The sampling rates to be considered are 100, 90, 80, $\dots$, 10\%. The 90\% credible intervals (CI) for the mean and variance parameters for all 100 datasets across all scenarios and sampling rates are then calculated. We assess the subsampling approach by comparing the empirical coverage, and determine the preferred data sampling rate as the one that results in the nominal coverage.

The results of the simulation study are provided in Appendix \ref{sec_appperc}. Overall, the correlation in the data affects the estimation of the emission distribution parameters, but the effect can be reduced by subsampling during the model fitting procedure. We use this result for the soccer referee example in Section \ref{sec:appl}. 

\section{Application to soccer referee data}
\label{sec:appl}

The HSMM specified in Equation \ref{eqcov} was used to model the heart rate of a referee during one game. The duration in each state is modeled using a zero-truncated Poisson distribution, where the duration parameter is defined as a function of three covariates. These include the average acceleration over the previous 20 seconds, the cumulative distance traveled during the game up to the time of transition, and the referee's distance from the center prior to the transition. These covariates capture the behavior of the referee over differing time scales which could have important impacts on heart rate, however other variations and covariates could be considered to study the demands of the game depending on the inferential goal. An indicator variable is also included to account for the difference between the first half (0) and second half (1), as well as an interaction between this indicator and each of the covariates to capture possible differing effects throughout the game. Including intercept terms, this results in 8 parameter coefficients for each state. Following the notation in Equation \ref{phif}, the duration parameter function in this application is defined as:

\begin{equation} 
\label{phifapp} 
\begin{gathered} 
	\phi_{S_{q+1}}(\mathbf{X}_{1:T_{q}},\bm \beta_{S_{q+1}}) = \exp \left [ \beta_{0,S_{q+1}} + \beta_{1,S_{q+1}} \widetilde{x}_{1,T_q} + \beta_{2,S_{q+1}} x_{2,T_q} + \beta_{3,S_{q+1}} x_{3, T_q} + \beta_{4,S_{q+1}} h_{T_{q+1}} \hspace{15mm} \right .\\ 
	\hspace{23mm} + \hspace{1mm} \left . \beta_{5,S_{q+1}} \left( \widetilde{x}_{1,T_q} h_{T_{q+1}} \right) + \beta_{6,S_{q+1}} \left( x_{2,T_q} h_{T_{q+1}} \right) + \beta_{7,S_{q+1}} \left( x_{3,T_q} h_{T_{q+1}} \right) \right] , 
\end{gathered} 
\end{equation} 

\noindent where $\widetilde{x}_{1,T_{q}} = \frac{1}{20} \sum\limits_{t=T_{q}-19}^{T_q} x_{1,t}$. That is, we consider the average over time points within 20 seconds of the transition in order to capture current movement and behavior of the referee.

One additional model specification is necessary since there is a break in play at half time. As such, an initial distribution was specified for the first state in the second half, similarly to the first state of the first half.

To determine the optimal subsampling rate for our analysis, we first ran the algorithm without subsampling and determined an approximate value for the size of the segments. The durations varied slightly by state and through time, but on average ranged from 29 to 39. We divided the data into segments to resemble the groups of emitted data by a state and calculated the mean autocorrelation across all groups. Considering group sizes of 29 to 39 resulted in autocorrelation values that ranged from 0.83 to 0.89, which we then compared to values in Table \ref{table_coveragesmu3s_4000}. Using $\psi=0.86$ suggests a subsampling rate of approximately 15\% for estimating the emission distribution parameters.

The number of states in HMMs and HSMMs has to be chosen {\it a priori} and it is usually determined through information criteria or expert knowledge \citep{liu_bayesian_2020}. However, the information criteria can lead to selecting a higher number of states given that they under-penalize model complexity, as mentioned in \cite{celeux_selecting_2008} and in the discussion on \cite{spiegelhalter_bayesian_2002}. In our case, the deviance information criterion (DIC) always directed us to selecting the model with the largest number of states, which led us to consider another measure for the selection. Specifically, we considered several chains for cases with 2, 3, 4 and 5 states, and observed that for 4 and 5 states the parameters did not converge to the same values. This suggested the use of a convergence diagnostic as an alternative to choose the appropriate number of states. The convergence was measured with the multivariate potential scale reduction factor (MPSRF) discussed in \cite{brooks_general_1998}. We selected a 3 state model as it was the only one with a MPSRF value less than or equal to 1.2. 

The MCMC algorithm was run for 100000 iterations. The first 30000 iterations were obtained using an adaptive random walk Metropolis algorithm for the duration distribution coefficients. These iterations were used to select the proposal variances for the random walk and then discarded. The remaining 70000 iterations were obtained based on these fixed proposal variances and these samples were used for parameter inference.

\begin{figure}[H]
	\begin{center}
		\includegraphics[scale=.6]{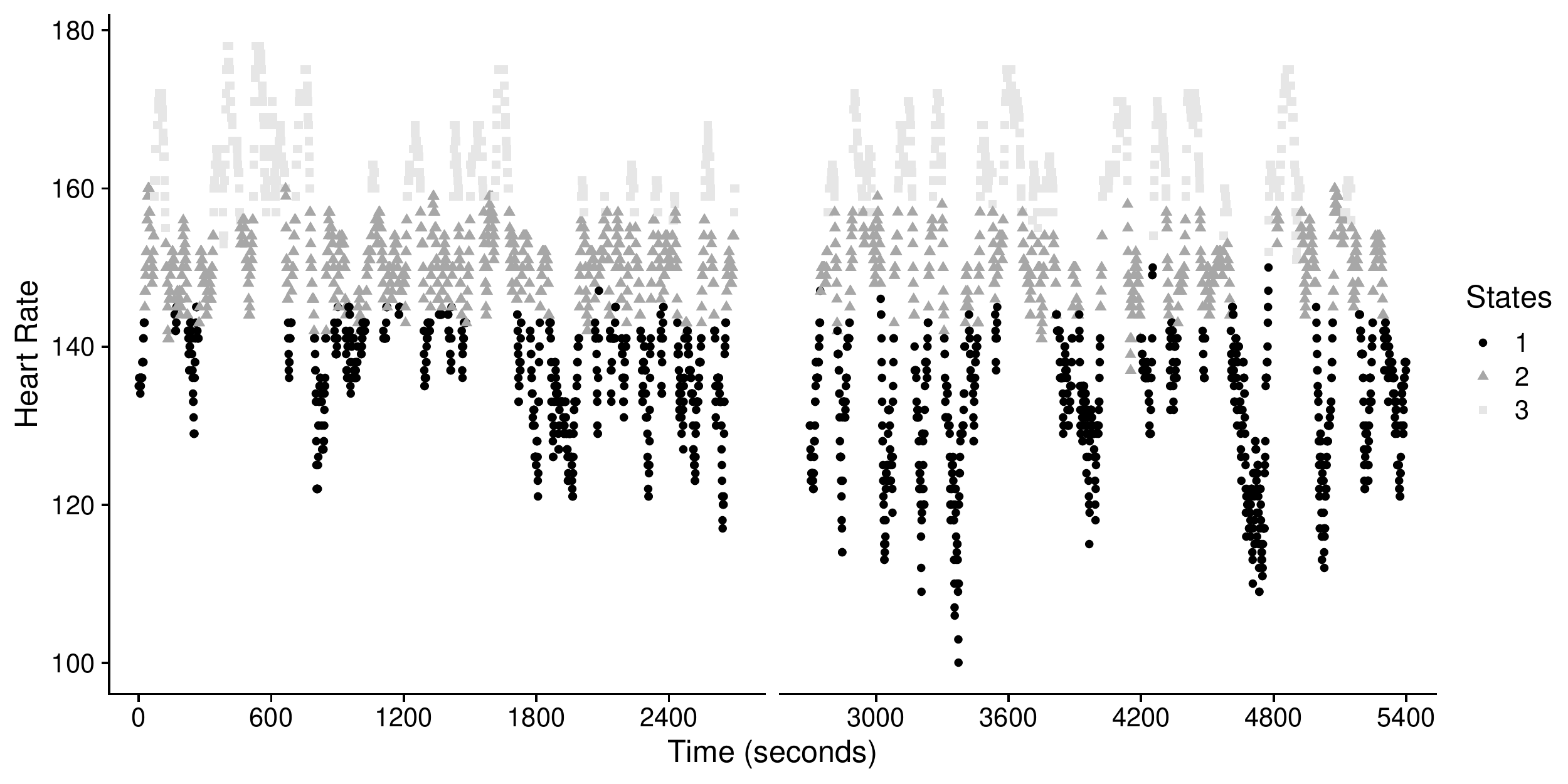}
		\caption{Referee's heart rate classified by latent states of their physiological response to movement during the game. \label{Fig:G3states}}
	\end{center}
\end{figure}

The posterior mean and 95\% credible intervals for the emission distribution parameters in each of the states is presented in Table \ref{table_meanG3}. There is a clear distinction between the mean heart rate in each state since none of the credible intervals overlap. The variability in the first state is notably higher, which can also be seen in Figure \ref{Fig:G3states} where the points corresponding to a lower heart rate, marked by circles, have a wider range. In general, heart rate is a physiological response to movement and is known to increase with increases in intensity of movement (e.g., walking versus running). In soccer, the referee's movement is dictated by the play of the game (e.g., ball in play, short passes, long passes, etc.). The first state can be thought to represent the referee's physiological response to movements of low or mild intensity, the second to moderate intensity, such as jogging, and the third to identify more strenuous intensity, such as running and sprinting. 

\begin{table}[H]
	\centering
	\caption{Posterior mean and 95\% CI of the emission distribution parameters}
	\label{table_meanG3}
	\begin{tabular}{lccc}
		\hline
		State & Mean & Variance \\
		\hline
S1 & 133.6 (132.3, 134.9) & 59.7 (47, 74.7) \\ 
S2 & 150.5 (149.8, 151.1) & 14.2 (10.9, 18.1) \\ 
S3 & 165 (164, 166) & 28.1 (22, 35.3) \\ 
		\hline
	\end{tabular}
\end{table}

The posterior probabilities of transitioning are shown in Table \ref{table_probG3}. Overall, the referee is more likely to transition between adjacent states (e.g., 1 to 2, or 2 to 3) than to jump from 1 to 3. This is also true for transitions from higher intensity states to lower intensity states (e.g., 3 to 2 and 2 to 1). This makes sense since heart rate observed at 1 second resolution will be relatively smooth. In addition, this is the result of the cardiac lag, where there is a known delay in the response time of increases (decreases) in heart rate in relation to increases (decreases) in the intensity of movement \citep{cort_cardiac_1991}. 

\begin{table}[H]
	\centering
	\caption{Posterior mean and 95\% CI for the state transition probabilities.}
	\label{table_probG3}
	\begin{tabular}{lcccc}
		\hline
		Transition & State 1 & State 2 & State 3  \\ 
		\hline
		State 1 & -- 				& 0.91 (0.81, 0.98)  & 0.09 (0.02, 0.19)\\ 
		State 2 & 0.56 (0.45, 0.68) & --				 & 0.44 (0.32, 0.55)\\ 
		State 3 & 0.03 (0, 0.12) 	& 0.97 (0.88, 1)  	& -- \\ 
		\hline
	\end{tabular}
\end{table}

Table \ref{table_durationratesG3} provides the number of segments in each state as well as summaries of the duration parameters for each state. 
The second state has more segments, yet the average duration parameter for this state is much smaller than the other two states.
The first and third state have fewer segments, yet both the mean and variation in the duration parameter is greater than for the second state. 
These differences in the duration parameters signify the importance of modeling the durations using covariates and state-specific parameters.

\begin{table}[H]
	\centering
	\caption{Segments and duration parameter statistics}
	\label{table_durationratesG3}
	\begin{tabular}{ccccc}
		\hline
		{\multirow{2}{*}{State}} &
		{\multirow{2}{*}{\# segments}} &
		\multicolumn{3}{c}{Duration parameter}\\
		\cline{3-5} 	
		\multicolumn{3}{c}{} Mean & Minimum & Maximum \\
		\hline
		S1 & 41 & 45.9 & 24.1 & 121 \\ 
		S2 & 69 & 27.8 & 11.6 & 70.4 \\ 
		S3 & 32 & 48.1 & 15.2 & 114.0 \\ 
		\hline
	\end{tabular}
\end{table}

The posterior means and credible intervals for the coefficients of the state-specific duration distribution parameters are given in Table \ref{table_betaG3v2}, with the significant coefficients presented in bold font. The coefficients of the duration distribution provide information about the average duration in the states as well as the difference in the covariate effects between the two halves.

\begin{figure}[H]
	\begin{center}
		\includegraphics[scale=.50]{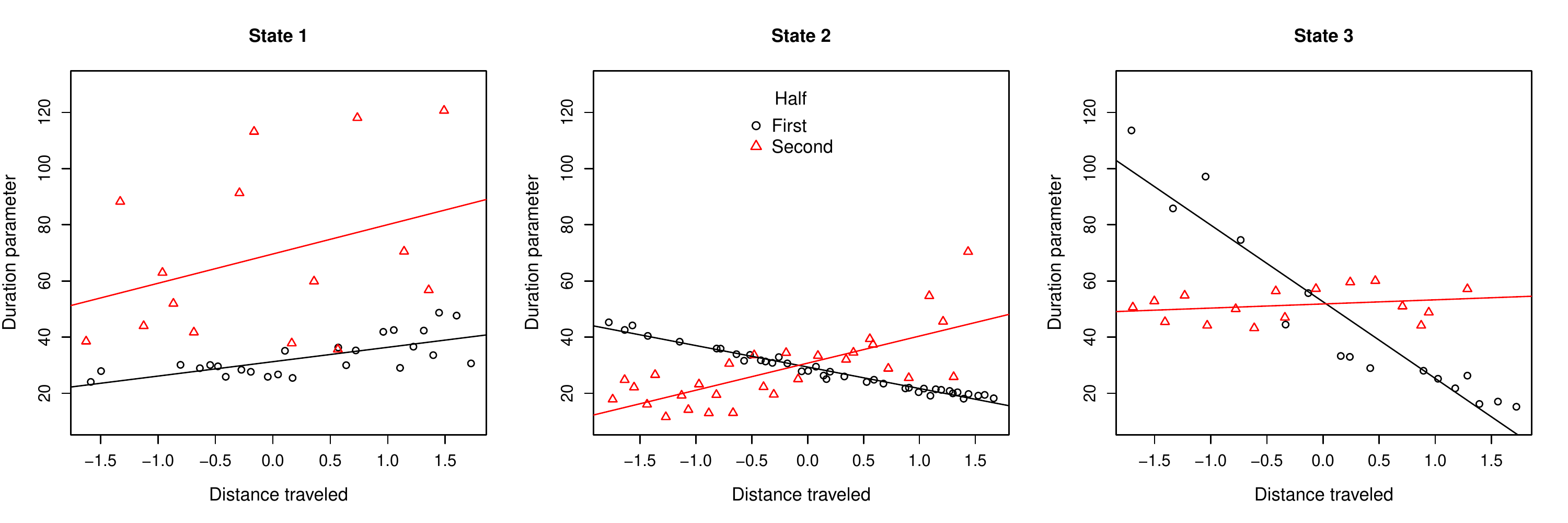}
		\caption{Scaled distance traveled and duration distribution parameter by state and half. \label{Fig:durationpar_covariates}}
	\end{center}
\end{figure}

In general, throughout the first half as the referee accumulated distance traveled, the duration of time in the first state increased, while the duration of time in second and third state decreased. 
In the second half, the duration of time in the first and second state increased as the referee accumulated distance traveled, while the duration of time in the third state remained constant.
These differences are depicted in Figure \ref{Fig:durationpar_covariates} for each half and state.
In the first half, transitions to the third state occurring after large deceleration events (negative acceleration) resulted in a longer duration in the state.
Transitions to the second state occurring after large acceleration events resulted in a shorter duration in the second half. We might conjecture that the referee's physiological response to higher intensity movements resulted in transitions from state one to state three with only a short duration in state two.
Lastly, when transitions happened further from the center during the first half, the duration of time in both the first and third state decreased. 
In the second half, when transitions happened further from the center, the duration of time increased for the first state and decreased for the second and third states.

\begin{table}[H]
	\centering
	\caption{Posterior mean and 95\% CI of the duration parameter coefficients.}
	\label{table_betaG3v2}
	\begin{adjustbox}{width=\columnwidth,center}
		\begin{tabular}{llccc}
			\hline
			Half & Variable & State 1 & State 2 & State 3 \\ 
			\hline
			{\multirow{4}{*}{1st }} &
			Intercept 					& \bf 3.42 (3.32, 3.51) & \bf 3.34 (3.24, 3.45) & \bf 3.79 (3.69, 3.89) \\ 
			{\multirow{4}{*}{}} &
			Acceleration (ACC) 			& -0.27 (-0.77, 0.29) & 0.09 (-0.39, 0.56) & \bf -0.74 (-1.32, -0.23) \\ 
			{\multirow{4}{*}{}} &
			Distance traveled (DT) 	 	& \bf 0.21 (0.12, 0.31) & \bf -0.27 (-0.36, -0.19) & \bf -0.64 (-0.74, -0.55) \\ 
			{\multirow{4}{*}{}} &
			Distance from center (DFC) 	& \bf -0.11 (-0.19, -0.03) & -0.03 (-0.12, 0.05) & \bf -0.09 (-0.17, -0.02) \\ 
			\hline
			{\multirow{4}{*}{2nd }} &
			Intercept					& \bf 4.17 (4.07, 4.27) & \bf 3.32 (3.18, 3.46) & \bf 3.92 (3.79, 4.03) \\ 
			{\multirow{4}{*}{}} &
			Acceleration (ACC)  		& 0.52 (-0.19, 1.15) & \bf -1.85 (-2.55, -1.18) & -0.13 (-0.5, 0.25) \\ 
			{\multirow{4}{*}{}} &
			Distance traveled (DT)  	& \bf 0.09 (0.02, 0.17) & \bf 0.34 (0.21, 0.49) & 0.03 (-0.09, 0.15) \\ 
			{\multirow{4}{*}{}} &
			Distance from center (DFC)  & \bf 0.34 (0.26, 0.42) & \bf -0.14 (-0.31, -0.02) & \bf -0.1 (-0.2, -0.01) \\ 
			\hline
		\end{tabular}
	\end{adjustbox}
\end{table}

\section{Discussion}
\label{sec:disc}

An extension of the HSMM was presented and applied to high-frequency wearable device data recorded for an athlete. We introduced non-homogeneity into the duration distribution of the HSMM using time-varying covariates.
In our example, the model enabled the characterization of a professional soccer referee's physiological response to movement. The three states for heart rate were characterized as a response to mild, moderate and strenuous activity, and their mean and variance were estimated.

Using a zero-truncated Poisson distribution for the duration in each state, we investigated the variation in time spent in each state as a function of important time-varying covariates. 
Important differences were detected in the relationship between the duration of time in each state and the covariates, and these differences varied between the first and second half.  
The variability of the duration parameters in the different segments supports the introduction of non-homogeneity in the HSMM for this application.

Analyzing athlete data from wearable devices with the proposed model provides information to athletes and trainers that can be used as part of performance evaluation and training session planning. Besides allowing activity recognition, our model informs about the length of time in the latent states and the factors related to this duration. Although not demonstrated here, this model could be applied to obtain predictions of the next states in the sequence and the amount of time an athlete would spend in those states given a set of prescribed movements.

Other approaches exist for introducing non-homogeneity into HSMMs. 
For example, parameters corresponding to transition probabilities could also be modeled as a function of covariates.
However, the focus of the analysis presented here was in capturing the variation in the duration of time in each state. 
Given that it is unlikely that the referee transitions from the state characterized as the result of low activity directly to that of high activity, it was not necessary to model the transition probabilities in terms of covariates for this application.

A further consideration for this application would be to include contextual information from the game. Given that the referee's behavior is dependent on the technical description of the game, variables related to passes or the length of time the ball is in play can be incorporated in the duration model. These covariates could also be used to model the transition probabilities or in the emission distribution.

The subsampling approach used in the estimation of the means and variances helped reduce the effect of dependence in the observation sequence. 
Our simulation study provided guidance as to the level of subsampling necessary to properly account for uncertainty in the model parameters. 
An indirect result of the subsampling was reduced computation time. 
Since high-frequency data are becoming increasingly common, additional research will focus on accounting for conditional dependence in the data. 

\bibliographystyle{apalike}
\bibliography{bhsmmref}

\newpage
\begin{appendix}

\counterwithin{table}{section}	
	
\section{Sampling algorithm}
\label{sec_appsa}

\begin{algorithm}
	\caption{MCMC sampling algorithm}
	\bigskip
	\textbf{Initial values}
	\begin{algorithmic}[1]
		\State Define initial values for parameters $\bm{\rho}^{(0)}, \textbf{P}^{(0)}, \textbf{B}^{(0)}$.
		\State Generate the state sequence $\textbf{S}^{(0)}$  as follows:
		\begin{algsubstates}
			\smallskip
			\State Sample the first state $S_{q=1}^{(0)}$ using $\bm{\rho}^{(0)}$.
			\smallskip
			\State Calculate $\phi_{S_{q=1}}^{(0)}$ using $\textbf{X}$ and $\bm{\beta}_{S_{q=1}}^{(0)}$.
			\smallskip
			\State Sample $\tau_1^{(0)}$ from a zero-truncated Poisson with parameter $\phi_{S_{q=1}}^{(0)}$.
			\medskip
			\State Define $\textbf{S}_{1:T_1}=S_{q=1}$.
			\smallskip
			\State Sample $S_{q=2}^{(0)}$ conditional on $S_{q=1}^{(0)}$ using $\textbf{P}^{(0)}$.
			\smallskip
			\State Calculate $\phi_{S_{q=2}}^{(0)}$ using $\textbf{X}$ and $\bm{\beta}_{S_{q=2}}^{(0)}$.
			\smallskip
			\State Sample $\tau_2^{(0)}$ from a zero-truncated Poisson with parameter $\phi_{S_{q=2}}^{(0)}$.
			\medskip
			\State Define $\textbf{S}_{{T_1}+1:T_2}=S_{q=2}$.
			\medskip
			\State Continue until $T_q=n$.
		\end{algsubstates}
		\State Calculate $\bm{\mu}^{(0)}$ and ${\bm{\sigma}^2}^{(0)}$ based on $\textbf{S}^{(0)}$.
	\end{algorithmic}
	\bigskip
	
	\textbf{Iterations}
	\begin{algorithmic}[1]
		\For{iteration $l=1,2,\dots$}
		
		\State Update $\bm{\rho}^{(l-1)}$ using Gibbs sampling:
		\begin{equation*}
			\bm{\rho}^{(l)} \sim Dir\left( I\left(S_{1}^{(l-1)}=1 \right)  + \theta_{\rho_1}, \dots, I\left(S_{1}^{(l-1)}=M \right)  + \theta_{\rho_M} \right) ,
		\end{equation*}
		where $I(\cdot)$ is the indicator function.	
		
		\State Update the $j$-th row of $\textbf{P}^{(l-1)}$ for $j=1, 2, \dots, M$, using Gibbs sampling:
		\begin{equation*}
			\textbf{P}_j^{(l)} \sim Dir\left( n_{j1}  + \theta_{P_{j1}}, \dots, n_{jM}  + \theta_{P_{jM}} \right) ,
		\end{equation*}
		where $n_{jk}= \sum\limits_{q=1}^{Q-1} I\left(S_{q}^{(l-1)}=j,S_{q+1}^{(l-1)}=k \right) $ is the total number of transitions from state $j$ to state $k$.

		\algstore{myalg}
	\end{algorithmic}
\end{algorithm}

\clearpage

\begin{algorithm}
	\ContinuedFloat
	\caption{MCMC sampling algorithm (continued)}
	\begin{algorithmic}
		\algrestore{myalg}
		
		\State Update $\bm{\beta}_j^{(l-1)}$ for $j=1, 2, \dots, M$ using random-walk Metropolis. Sample $\textbf{z}\sim N(\textbf{0},\kappa_{\beta_j}^2 \textbf{I}_{r+1})$, where $ I_{r+1}$ is an identity matrix of order $r+1$, and define the proposal vector $\bm{\beta}_j^{(*)}=\bm{\beta}_j^{(l-1)} + \textbf{z}$. Then calculate the Metropolis ratio as:
		\begin{equation*}
			m_{\bm{\beta}_j}= \left( \frac{ \prod \limits_{q=1}^Q ZTP \left(\tau_{q}^{(l-1)} \mid \phi_q^{(*)} \right)}{\prod \limits_{q=1}^Q ZTP \left(\tau_{q}^{(l-1)} \mid \phi_q^{(l-1)} \right)} \right) \times \left( 	\frac{N\left(\bm{\beta}_j^{(*)} \mid \bm{\theta}_{\beta_j}, \lambda_{\beta_j}^2 \textbf{I}_{r+1} \right) }{N\left(\bm{\beta}_j^{(l-1)} \mid \bm{\theta}_{\beta_j}, \lambda_{\beta_j}^2 \textbf{I}_{r+1} \right) } \right)
			,	
		\end{equation*}
		and if $u < m_{\bm{\beta}_j}$, with $u\sim Unif(0,1)$, let  $\bm{\beta}_j^{(l)}=\bm{\beta}_j^{(*)}$, and update $\bm\phi^{(l-1)}$, $\bm\phi^{(l)}=\bm\phi^{(*)}$.
		
		\State Update $\textbf{S}^{(l-1)}$ with the Viterbi algorithm in \cite{yu_hidden_2016} and then update $\bm\tau^{(l-1)}$. 
		
		\State Subsample the observations. Randomly sample the observations $\textbf{y}$ according to the specified sampling rate, and use the new observation vector $\tilde{\textbf{y}}$ of size $\tilde{n}$ in steps 10 and 11.	
		
		\State Update $\mu_j^{(l-1)}$ for $j=1, 2, \dots, M$ using Gibbs sampling and the subsampling approach: 
		\begin{equation*}
			\mu_j \sim {N} \left(  \frac{\frac{\sum \limits_{i=1}^{n} y_i \cdot {I({S}_i=j, y_i \in \tilde{\textbf{y}})}}{{\sigma_{j}^2}^{(l-1)}} + \frac{\theta_\mu}{\lambda_\mu^2}}{ \frac{\tilde{n}_{j}}{{\sigma_{j}^2}^{(l-1)}} + \frac{1}{\lambda_\mu^2}}, \frac{1}{ \frac{\tilde{n}_{j}}{{\sigma_{j}^2}^{(l-1)}} + \frac{1}{\lambda_\mu^2}} \right),
		\end{equation*}
		where $I(\cdot)$ is the indicator function and $\tilde{n}_j$ is the total number of observations of vector $\tilde{\textbf{y}}$ emitted by state $j$.
		
		\State Update ${\sigma_j^2}^{(l-1)}$ for $j=1, 2, \dots, M$  using Gibbs sampling and the subsampling approach:
		\begin{equation*}
			\sigma_{j}^2 \sim IG \left( \theta_{\sigma^2} + \frac{\tilde{n}_j}{2} ,   \lambda_{\sigma^2} + \frac{1}{2}\sum \limits_{i=1}^{n} \left( y_i \cdot {I({S}_i=j, y_i \in \tilde{\textbf{y}})} - \mu_j^{(l-1)} \right) ^2   \right) .
		\end{equation*}
		
		\State Save $\bm{\rho}^{(l)}, \textbf{P}^{(l)}, \textbf{B}^{(l)}, \bm\phi^{(l)},\textbf{S}^{(l)},\bm\tau^{(l)}, \bm{\mu}^{(l)}$ and ${\bm{\sigma}^2}^{(l)}$.
		
		\EndFor
	\end{algorithmic}
\end{algorithm}

\section{Simulation study results}
\label{sec_appperc}

The simulation study was developed for 2, 3 and 4 states. The empirical coverage is calculated as the percentage of 90\% CIs that captured the true parameter values. Tables \ref{table_coveragesmu2s_4000} to \ref{table_coveragesmu4s_4000} show to the case where sample size was approximately 4000, and they present the empirical coverage for the different autocorrelation parameters and sampling rates utilized. This coverage corresponds to the average coverage of the different state means. The empirical coverage for the scenarios with sample sizes 1000 and 2600 was similar to the case with sample size 4000. 

There are two important results provided in the tables. First, the more correlated the data are, the worse we do in recovering the true parameter values, as indicated by the first row of the tables where no subsampling was used. Second, the empirical coverage increases as the sampling rate decreases. As we reduce the percent of data used, we are able to reduce the dependence in the data and thus improve our estimates of uncertainty. For a case where the autocorrelation is approximately 0.86 and the whole dataset is utilized (sampling rate = 100\%), then the coverage is low, indicating the need to use a smaller sampling rate. With data having autocorrelation close to 0.3, we obtain nominal coverage of the emission distribution means when using only 70\% of the data points in the MCMC iterations.

\begin{table}[H]
	\centering
	\caption{Coverage percentage of the emission distribution means, 2 states case with n$\approx$4000.}
	\label{table_coveragesmu2s_4000}
	\begin{tabular}{ccccc}
		\hline 
		{\multirow{2}{*}{Sampling rate \%  }} &
		\multicolumn{4}{c}{$\psi$}\\
		\cline{2-5} 	
		\multicolumn{2}{c}{} 0.30 & 0.60 & 0.86 & 0.95\\
		\hline
		100 & 76 & 66 & 36 & 25 \\ 
		90 & 80 & 70 & 40 & 27 \\ 
		80 & 85 & 75 & 46 & 32 \\ 
		70 & 88 & 78 & 51 & 36 \\ 
		60 & 92 & 83 & 56 & 39 \\ 
		50 & 94 & 88 & 62 & 46 \\ 
		40 & 99 & 93 & 70 & 54 \\ 
		30 & 100 & 99 & 75 & 62 \\ 
		20 & 100 & 100 & 86 & 76 \\ 
		10 & 100 & 100 & 96 & 91 \\ 
		\hline
	\end{tabular}
\end{table}

\begin{table}[H]
	\centering
	\caption{Coverage percentage of the emission distribution means, 3 states case with n$\approx$4000.}
	\label{table_coveragesmu3s_4000}
	\begin{tabular}{ccccc}
		\hline 
		{\multirow{2}{*}{Sampling rate \%  }} &
		\multicolumn{4}{c}{$\psi$}\\
		\cline{2-5} 	
		\multicolumn{2}{c}{} 0.30 & 0.60 & 0.86 & 0.95\\
		\hline
		100 & 78 & 57 & 35 & 24 \\ 
		90 & 83 & 62 & 39 & 24 \\ 
		80 & 86 & 67 & 42 & 28 \\ 
		70 & 91 & 71 & 46 & 33 \\ 
		60 & 94 & 78 & 51 & 35 \\ 
		50 & 96 & 82 & 57 & 42 \\ 
		40 & 98 & 90 & 64 & 46 \\ 
		30 & 100 & 95 & 69 & 53 \\ 
		20 & 100 & 98 & 82 & 63 \\ 
		10 & 100 & 100 & 94 & 84 \\ 
		\hline
	\end{tabular}
\end{table}

\begin{table}[H]
	\centering
	\caption{Coverage percentage of the emission distribution means, 4 states case with n$\approx$4000.}
	\label{table_coveragesmu4s_4000}
	\begin{tabular}{ccccc}
		\hline 
		{\multirow{2}{*}{Sampling rate \%  }} &
		\multicolumn{4}{c}{$\psi$}\\
		\cline{2-5} 	
		\multicolumn{2}{c}{} 0.30 & 0.60 & 0.86 & 0.95\\
		\hline
		100 & 78 & 62 & 38 & 28 \\ 
		90 & 83 & 66 & 40 & 31 \\ 
		80 & 87 & 72 & 43 & 35 \\ 
		70 & 91 & 78 & 46 & 38 \\ 
		60 & 93 & 83 & 53 & 42 \\ 
		50 & 96 & 87 & 60 & 46 \\ 
		40 & 97 & 90 & 65 & 53 \\ 
		30 & 99 & 96 & 74 & 62 \\ 
		20 & 100 & 98 & 85 & 74 \\ 
		10 & 100 & 100 & 96 & 88 \\ 
		\hline
	\end{tabular}
\end{table}

\end{appendix}

\end{document}